\documentstyle[epsf]{mn}

\def\apj{ApJ}
\def\mnras{MNRAS}
\def\apjs{ApJS}
\def\apjl{ApJ}
\def\aj{AJ}

\def\aap{Astr.\ Astrophys.}

\title [A wide-field {\em K}--band survey -- I. Galaxy counts]
{A wide-field {\em K}--band survey -- I. Galaxy counts in {\em B},
{\em V}, {\em I} and {\em K}}

\author[J. P. Gardner et al.] {J. P. Gardner$^{1,3}$, R. M.
Sharples$^{1}$, B. E. Carrasco$^{2,3}$, \& C. S. Frenk$^{1}$
\\$^1$University of Durham, Department of Physics, South Road,
Durham DH1 3LE \\ $^2$INAOE, Apdo Postal 216 y 51, Puebla, CP 72000,
MEXICO \\ $^3$Visiting Astronomer, Kitt Peak National Observatory,
National Optical Astronomy Observatories, operated by AURA Inc., \\
under contract with the National Science Foundation, USA. \\ E-mail
addresses: jonathan.gardner@durham.ac.uk; r.m.sharples@durham.ac.uk;
bec@tonali.inaoep.mx; c.s.frenk@durham.ac.uk}

\date{Accepted for publication in MNRAS}

\begin{document}

\maketitle

\begin{abstract}

We present bright galaxy number counts measured with linear detectors
in the $B$, $V$, $I$, and $K$ bands in two fields covering nearly
10 square degrees. All of our measurements are consistent with
passive evolution models, and do not confirm the steep slope measured
in other surveys at bright magnitudes. Throughout the range $16 <
B < 19$, our $B-$band counts are consistent with the ``high
normalization'' models proposed to reduce the faint blue galaxy
problem. Our $K-$band counts agree with previous measurements, and
have reached a fair sample of the universe in the magnitude range
where evolution and K-corrections are well understood.

\end{abstract}


\section{Introduction}

Observational studies of galaxy formation and evolution have
advanced at an unprecedented pace in recent years. Two developments
have played a key role: CCD imagery to very faint limits and the
ability to measure redshifts for large samples at increasingly
faint magnitudes. Progress to date has relied primarily on optical
data, but it has been clear for some time that near-infrared
observations are fundamental. Samples selected according to $K-$band
flux are superior to the traditional $B-$selected samples in at
least three respects: {\it (i)} In the infrared, K-corrections due
to the redshift of the spectral energy distribution are smooth,
well-understood, and nearly independent of Hubble type; the expected
luminosity evolution is also smooth. {\em (ii)} At high redshift,
the observer's near-infrared samples the well-understood rest frame
optical (dominated by long-lived, near-solar mass stars), while
the optical band samples the poorly-understood rest frame ultraviolet
(dominated by short-lived massive stars.) {\em (iii)} Since
near-solar mass stars make up the bulk of a galaxy, the absolute
$K$ magnitude is a measure of the visible mass in a galaxy. 

Deep photometric surveys of small areas measuring the number counts
and colours of field galaxies have reached as faint as $K=24$,
(Gardner, Cowie \& Wainscoat 1993; Cowie et al.\ 1994; Djorgovski
et al.\ 1995,) and show significant amounts of galaxy evolution at
high redshift. However, the interpretation of faint galaxy data
(either counts or redshift distributions) hinges on an accurate
statistical description of the local population of galaxies.
Photometric and spectroscopic surveys of bright galaxies, covering
an area large enough to average over the effects of large-scale
structure, are required in order to obtain bright galaxy counts
and colour distributions, measurements of the galaxy-galaxy
correlation function, and the local luminosity function. In the
$K$ band, the small size of infrared detectors and the corresponding
small field of view available, has made it difficult in the past
to image large areas. In the optical, much work has been done using
photographic plates, but their non-linearity can introduce possible
systematics in the photometry (Metcalfe, Fong \& Shanks 1995.)

We have imaged nearly 10 square degrees in two fields at high
galactic latitude using a NICMOS3 detector in the near-infrared
$K$ band, and a CCD camera in the optical $B$, $V$, and $I$ bands.
Here, we present the galaxy number counts. In a companion paper
(Baugh et al.\ 1996, Paper II) we present the galaxy correlation
function. Our $K<15$ photometric catalog has been used to select
galaxies for spectroscopic follow-up, and future papers in this
series will present the $K-$band galaxy luminosity function, the
galaxy redshift and colour distributions, and a discussion of the
star counts and colour distribution. One of our fields, centered
on the north ecliptic pole, is ideally situated for viewing by
satellites in polar orbits, and has been the subject of deep IRAS
and ROSAT observations.

\section{The Data}

We have imaged 9.84 square degrees in the $K-$band with a $256^2$
HgCdTe NICMOS3 detector, and in the $B$, $V$, and $I$ bands with
a $2048^2$ CCD camera. The $K-$band observations were made in 1994
June with the IRIM camera on the Kitt Peak National Observatory
$1.3 m$ telescope. On this telescope, the IRIM camera has $1.96^{\prime
\prime}$ pixels and an $8.36^{\prime}$ field of view. Each point
was observed with at least two $60 sec$ exposures, reaching a
$5\sigma$ galaxy detection depth of $K=15.6$ in a $10^{\prime
\prime}$ circular aperture. The $3\sigma$ surface brightness limit
of our images is $K=18.5 ~ mag ~ arcsec^{-2}$. The $B-$, $V-$, and
$I-$band observations were made in 1995 June with the T2KA camera
on the KPNO $0.9 m$ telescope. On this telescope, the T2KA CCD has
$0.68^{\prime \prime}$ pixels with a $23.2^{\prime}$ field of view.
Each point in each filter was observed with a $300 sec$ exposure,
and the images reach a $5 \sigma$ detection depth of $B=21.1$,
$V=20.9$, and $I=19.6$ in a $10^{\prime \prime}$ circular aperture.
The $3\sigma$ surface brightness limit is $24.0$, $23.8$, and $22.5
~ mag ~ arcsec^{-2}$ for $B$, $V$ and $I$ respectively. The location
of the two fields were selected randomly, that is, without regard
to the presence or absence of any known objects. The field centers
are at RA $14^h 15^m$, Dec $+00$ and RA $18^h 00^m$, Dec $+66$;
galactic latitudes $+55$ and $+30$ respectively. One of our fields
has a nearby rich galaxy cluster within it, and to avoid biasing
the galaxy counts, we have removed all objects within 1 degree
radius of the central galaxy of this cluster. Thus the effective
area for the counts presented here is 8.54 square degrees.

\begin{table}
\caption{The galaxy number counts}
\label{nctable}

\begin{tabular}{@{}lrrrrr}
Filter & Mag & Raw N &  $log(N)$ & ${\sigma}_{high}$ & ${\sigma}_{low}$ \\
$K$ &  10.25  &    1 &  -0.630   &   0.519   &   0.762  \\
    &  10.75  &    1 &  -0.630   &   0.519   &   0.762  \\
    &  11.25  &    4 &  -0.028   &   0.253   &   0.283  \\
    &  11.75  &   13 &   0.484   &   0.134   &   0.139  \\
    &  12.25  &   22 &   0.712   &   0.101   &   0.103  \\
    &  12.75  &   33 &   0.888   &   0.082   &   0.083  \\
    &  13.25  &   66 &   1.189   &   0.057   &   0.057  \\
    &  13.75  &  138 &   1.510   &   0.039   &   0.039  \\
    &  14.25  &  273 &   1.806   &   0.027   &   0.027  \\
    &  14.75  &  642 &   2.177   &   0.018   &   0.018  \\
    &  15.25  & 1290 &   2.480   &   0.012   &   0.012  \\
    &  15.75  & 2609 &   2.786   &   0.009   &   0.009  \\
    &         &      &           &           &          \\
$I$ &  12.25  &    2 &  -0.329   &   0.365   &   0.451  \\
    &  12.75  &    1 &  -0.630   &   0.519   &   0.762  \\
    &  13.25  &    6 &   0.148   &   0.203   &   0.219  \\
    &  13.75  &   11 &   0.411   &   0.147   &   0.153  \\
    &  14.25  &   23 &   0.731   &   0.099   &   0.101  \\
    &  14.75  &   26 &   0.785   &   0.093   &   0.094  \\
    &  15.25  &   63 &   1.169   &   0.058   &   0.058  \\
    &  15.75  &   95 &   1.347   &   0.047   &   0.047  \\
    &  16.25  &  198 &   1.666   &   0.032   &   0.032  \\
    &  16.75  &  398 &   1.970   &   0.022   &   0.022  \\
    &  17.25  &  644 &   2.179   &   0.017   &   0.018  \\
    &  17.75  & 1190 &   2.445   &   0.013   &   0.013  \\
    &         &      &           &           &          \\
$V$ &  12.25  &    1 &  -0.630   &   0.519   &   0.762  \\
    &  13.25  &    1 &  -0.630   &   0.519   &   0.762  \\
    &  13.75  &    1 &  -0.630   &   0.519   &   0.762  \\
    &  14.25  &    5 &   0.069   &   0.224   &   0.246  \\
    &  14.75  &    6 &   0.148   &   0.203   &   0.219  \\
    &  15.25  &   14 &   0.516   &   0.129   &   0.133  \\
    &  15.75  &   25 &   0.768   &   0.095   &   0.096  \\
    &  16.25  &   48 &   1.051   &   0.067   &   0.067  \\
    &  16.75  &   83 &   1.289   &   0.050   &   0.050  \\
    &  17.25  &  142 &   1.522   &   0.038   &   0.038  \\
    &  17.75  &  285 &   1.824   &   0.026   &   0.027  \\
    &  18.25  &  454 &   2.027   &   0.021   &   0.021  \\
    &  18.75  &  766 &   2.254   &   0.016   &   0.016  \\
    &         &      &           &           &          \\
$B$ &  12.25  &    1 &  -0.630   &   0.519   &   0.762  \\
    &  14.25  &    3 &  -0.153   &   0.295   &   0.341  \\
    &  14.75  &    2 &  -0.329   &   0.365   &   0.451  \\
    &  15.25  &    6 &   0.148   &   0.203   &   0.219  \\
    &  15.75  &    6 &   0.148   &   0.203   &   0.219  \\
    &  16.25  &   20 &   0.671   &   0.106   &   0.109  \\
    &  16.75  &   26 &   0.785   &   0.093   &   0.094  \\
    &  17.25  &   56 &   1.118   &   0.062   &   0.062  \\
    &  17.75  &  117 &   1.438   &   0.042   &   0.042  \\
    &  18.25  &  188 &   1.644   &   0.033   &   0.033  \\
    &  18.75  &  323 &   1.879   &   0.025   &   0.025  \\
    &  19.25  &  509 &   2.076   &   0.020   &   0.020  \\
    &  19.75  &  901 &   2.324   &   0.015   &   0.015  \\
    &         &      &           &           &          \\
\end{tabular}

The raw number of galaxies in the 8.54 square degree area, and the
$log(N/mag/deg^2)$. The high and low Poissonian errors are taken
from the calculations of Gehrels (1986).

\end{table}

\begin{figure}

\epsfxsize=8.5cm

\epsfysize=8.5cm

\epsfbox{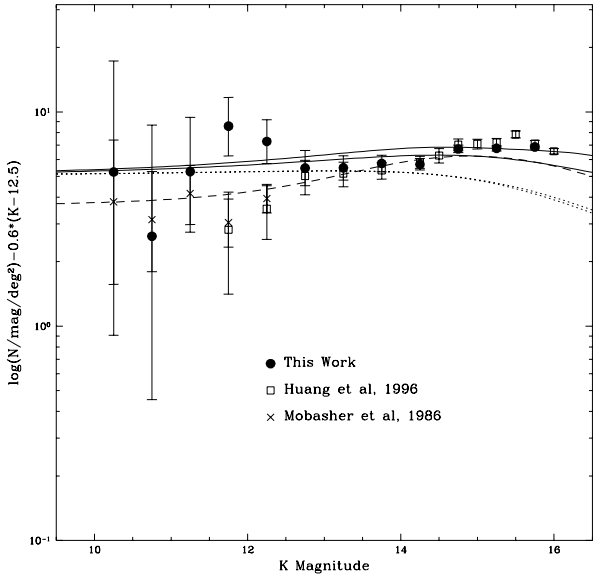}

\caption{The $K-$band galaxy counts. The Euclidean slope, $d
log(n)/dm = 0.6$, has been subtracted in order to expand the
ordinate. The models plotted are no-evolution (dotted line) and
passive-evolution (solid line), in an open universe (the two middle
lines at faint magnitudes) and in a closed universe. The dashed
line is the no-evolution, local underdensity model of Huang et
al.\ (1996).}

\label{knm}

\end{figure}

Image reduction will be described in detail elsewhere (Gardner et
al, in preparation), but mainly followed the techniques described
by Gardner (1995). Briefly, the $K-$band images were dark-subtracted,
flattened with domeflats, median sky-subtracted, and large-scale
gradients were removed with a large ($2^{\prime}$) median filter.
The optical images were bias subtracted, and flattened with twilight
flats. In addition, the $I-$band images were flattened with median
sky flats. Object identification on the optical images was done
with the SExtractor program (Bertin \& Arnouts 1996), using a
$3\sigma$ threshold. Deblending was done with a multi-thresholding
approach, and integral pixels were assigned to each object. Magnitudes
are the SExtractor {\em mag\_best}, which is the flux in an elliptical
aperture at $2.5$ times the Kron (1980) radius along the semi-major
and semi-minor axes for isolated objects, and a corrected-isophotal
magnitude when deblending was required. $K-$band photometry was
done in $10^{\prime \prime}$ apertures on all identified $I-$band
objects. The aperture magnitudes were deblended by assigning pixels
by hand, and were corrected to total magnitudes using the curve of
growth measured in the $I$ band. Colours for all objects were
measured within $5^{\prime \prime}$ circular apertures in the
optical, and within $6^{\prime \prime}$ square apertures in $I-K$.

Star/galaxy separation is critical at these magnitudes, and was
carried out using a combination of morphology on the optical images,
and the colour criterion discussed by Gardner (1995). The seeing
of the optical images varied through the observing run between two
and three pixels, or $1.3^{\prime \prime} < fwhm < 2.0^{\prime
\prime}$. All identified galaxies with $K<15$, $I<18$, $V<19$ or
$B<20$ were confirmed by eye. The median Kron (1980) $r_{1}$ radius
in the faintest magnitude bin on the optical images was typically
$2.5^{\prime \prime}$ for stars and $3.4^{\prime \prime}$ for
galaxies. The $I-K$ colour, in combination with the $B-I$ colour,
is a good indicator of star/galaxy separation for all but the bluest
objects. We found no large population of compact objects with the
colours of galaxies, nor did we find a large population of extended
objects with the colours of stars. In the range $15<K<16$, star/galaxy
separation was carried out on the basis of colour alone, using
$V-I$ vs $I-K$ for the objects not detected in $B$.

The galaxy number counts are presented in Table~\ref{nctable}, and
the $K-$band counts are plotted in Figure~\ref{knm}. To expand the
ordinate, we have subtracted the Euclidean slope $d log(n)/ dm =
0.6$. Alongside our data, we plot other existing bright $K-$band
galaxy counts. For comparison we also show the predictions of a
simple model, based upon the formulation of Yoshii \& Takahara
(1988), modified to include rest-frame and evolved spectral energy
distributions from the GISSEL models (Bruzual \& Charlot 1993; 1996
in preparation.) The model is similar to that used in Gardner
(1996), except that we have used the revised version of the GISSEL
models. For the current purposes, the main difference between the
two versions is that rest frame optical-near infrared colours are
redder.  The solid lines include this passive evolution, while the
dotted lines are no-evolution models, i.e. models that include only
the cosmological geometry and K-corrections. The dashed line is a
``local-underdensity'' model proposed by Huang et al.\ (1996) and
is discussed below. To constuct the models we adopted the $b_J$
type-independent luminosity function of Loveday et al.\ (1992),
converted to type-dependent luminosity functions in other filters
through rest frame colours. The normalization of the models was
determined with a least-squares fit of the passive-evolution flat
universe model to our data.

\begin{figure}

\epsfxsize=8.5cm

\epsfysize=8.5cm

\epsfbox{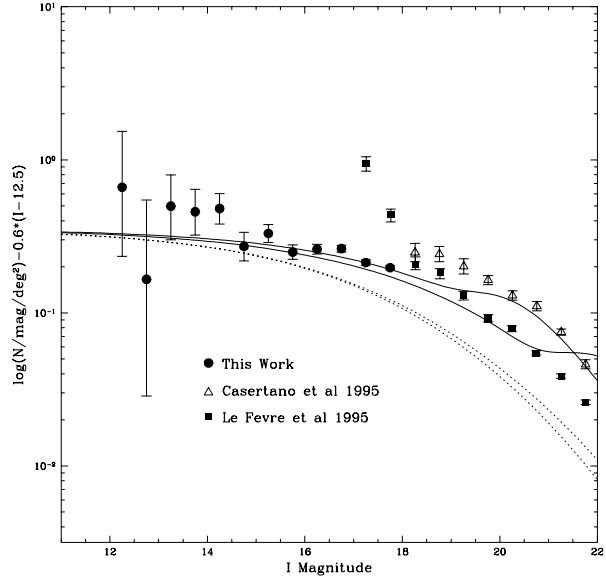}

\caption{The $I-$band number counts. The models plotted are as in
Figure~\protect\ref{knm}. All counts have been converted to $I_{kc}$ 
as discussed in the text. The Medium Deep Survey counts of
Casertano et al.\ (1995) show an excess over our counts and over those
of the CFRS (Le Fevre et al.\ 1995).}

\label{inm}

\end{figure}

Figure~\ref{inm} shows our $I-$band galaxy counts, again with the
Euclidean slope subtracted, together with other existing counts
converted to the Kron-Cousins $I$ band which we used. Figure~\ref{vnm}
shows our $V-$band galaxy counts and Figure~\ref{bnm} our $B-$band
galaxy counts.

\section{Discussion}

The $K-$band galaxy counts presented in Table~\ref{nctable} and
Figure~\ref{knm} show remarkable consistency with the counts of
Huang et al.\ (1996), which were also based upon $K-$band observations
of approximately 10 square degrees. While one of us (JPG) is also
a co-author of that paper, the two data sets were collected, reduced
and analyzed independently, and there is no overlap in area between
the two surveys. The consistency of the two results indicates that
both surveys represent a fair sample of the universe, so the bright
$K-$band galaxy counts and their normalization are no longer a
subject of debate.

Our counts in the region $13 < K < 15$ are within $1 \sigma$ of
the Huang et al.\ (1996) counts in each magnitude bin. Nevertheless,
we measure a shallower slope. The slope of our counts is $0.627
\pm 0.010$, consistent with the flat-universe passive evolution
model plotted in Figure~\ref{knm}. Thus our data do not support
the conclusion of Huang et al.\ (1996) that a model with a steep
slope is required to fit the bright $K-$band number counts. The
value of the normalization, $\phi^* = 1.50 \times 10 ^{-2} h^3 Mpc
^{-3}$, (for $H_0=100h~km~s^{-1}~Mpc^{-1}$,) inferred from our
counts is higher than the value in the Loveday et al.\ (1992)
luminosity function converted to the $K-$band. This comparison,
however, is uncertain because the transformation from the $b_J$-band
luminosity function to a $K-$band luminosity function is very model
dependent. The model $K^*$ depends strongly on the assumed rest-frame
colours and the 3 parameters of the Schechter (1976) luminosity
function are strongly correlated. Neither of the two existing
measurements of the $K-$band luminosity function surveyed enough
galaxies to accurately constrain $\phi^*$ (Mobasher, Sharples \&
Ellis 1993; Glazebrook et al.\ 1995). In the absence of an accurate
measurement of the joint $[B,K]$ luminosity function, it is difficult
to determine the consistency of the normalization of number count
models in different filters. For this reason, we have normalized
the models plotted in each of the figures to our data.

To fit the bright $K-$band number counts without passive evolution,
Huang et al.\ (1996) have proposed that the local universe is
underdense by a factor of 2. They have constructed a heuristic
model in which this underdensity smoothly dampens out by $z \approx
0.2$. This model is unphysical and, as they stress, the number
counts alone do not contain enough information to constrain its
specifics. Maddox et al.\ (1990) measured a steep slope in the
bright $B-$band number counts and a local underdensity has also
been proposed to explain this slope (Shanks 1990; Metcalfe et
al.\ 1991). We plot the model of Huang et al.\ (1996) in Figure~\ref{knm}
(with their normalization) as a dashed line. This model does not
fit our galaxy counts.

The $I-$band galaxy counts are presented in Table~\ref{nctable}
and Figure~\ref{inm}. We used the Kitt Peak standard $I$ filter,
corrected to the Kron-Cousins standard stars of Landolt, (1983;
1992). Also plotted in Figure~\ref{inm} is a compilation of $I-$band
counts, converted to Kron-Cousins $I_{kc}$, taken from the
Canada-France Redshift Survey (Le Fevre et al.\ 1995, hereafter
CFRS) and the WF/PC Medium Deep Survey counts of Casertano et
al.\ (1995), (hereafter MDS). The latter were converted  using
$I_{kc} = I_{785} + 0.16* (V_{555} - I_{785})$ and $(V_{555} -
I_{785})=1.6$ (Edvardsson \& Bell 1989). While there are no other
measurements of the $I-$band number counts as bright as ours, the
faint end of our counts are consistent with the CFRS counts. They
are not, however, consistent with the MDS counts, which are a factor
of 1.5 higher. The MDS measurements are based upon pre-refurbishment
WF/PC images. Casertano et al.\ (1995) found that the median
half-light radius of their galaxies was $0.3^{\prime \prime}$, so
star/galaxy separation based upon morphology alone would be very
difficult. They ascribe the excess in their number counts to
inaccurate star/galaxy separation in the ground-based data. However,
as noted above, our star/galaxy separation is based upon morphology
and colour, and we do not find a large population of compact objects
with the colours of galaxies. In addition, the CFRS collaboration
obtained spectroscopy of one sixth of the objects in their sample,
without regard to morphology, and also did not find a large population
of compact galaxies. The excess in the MDS counts is most likely
due to errors in the photometry of the WF/PC images introduced in
the deconvolution process.

\begin{figure}

\epsfxsize=8.5cm

\epsfysize=8.5cm

\epsfbox{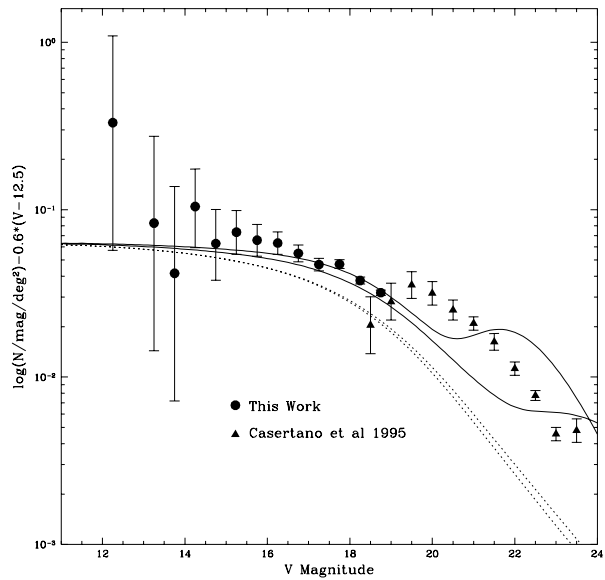}

\caption{The $V-$band number counts. The models plotted are as in
Figure~\protect\ref{knm}.}

\label{vnm}

\end{figure}

The $V-$band galaxy counts are presented in Table~\ref{nctable}
and Figure~\ref{vnm}. We used the Kitt Peak standard $V$ filter,
corrected to the Johnson $V$ standard stars of Landolt, (1983;
1992). There are only two other measurements of the $V-$band number
counts, by Casertano et al.\ (1995) (MDS), and by Driver et
al.\ (1994). $V_{555}$ is approximately equal to Johnson $V$ for
the galaxies in these samples, so we have applied no correction.
The MDS $V$ counts again show an excess over the faint end of the
counts presented here.

\begin{figure}

\epsfxsize=8.5cm

\epsfysize=8.5cm

\epsfbox{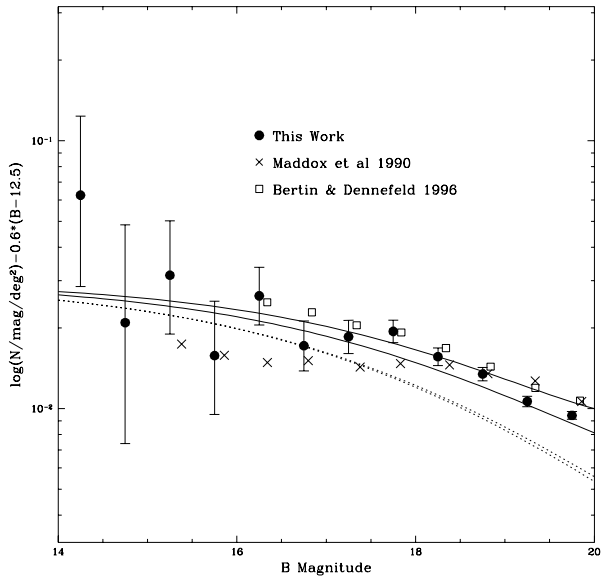}

\caption{The $B-$band number counts. The models plotted are as in
Figure~\protect\ref{knm}. All counts have been converted to Johnson
$B$ as discussed in the text. Our counts show a shallower slope
than the APM counts of Maddox et al.\ (1990), and are consistent with
passive evolution models.}

\label{bnm}

\end{figure}

The $B-$band galaxy counts are presented in Table~\ref{nctable}
and Figure~\ref{bnm}. We used the Kitt Peak standard $B$ filter,
corrected to the Johnson $B$ standard stars of Landolt, (1983;
1992). The $B$ counts have been measured in many surveys, but this
is the first time that $B$ counts at $B<20$ have been obtained with
a CCD camera, rather than with non-linear photographic plates.
While our area is much smaller than other surveys, and our statistical
error is higher, our counts are far less likely to suffer from
systematic effects in the photometry. In Figure~\ref{bnm} we have
plotted $b_J$ counts, converted to the Johnson $B-$band, from the
APM survey of Maddox et al.\ (1990) (hereafter APM), and from the
MAMA survey of Bertin \& Dennefeld (1996).

The APM counts show the steep slope at the bright end mentioned
earlier. These authors interpretated the data as revealing a large
(and unexpected) amount of luminosity evolution at low redshifts
($z<0.1$). Other workers have attributed this steep slope to a
local underdensity of galaxies (Shanks 1990; Metcalfe et al.\ 1991),
to a selection effect against low surface brightness galaxies
(McGaugh 1994; Ferguson \& McGaugh 1995), or to systematics in the
photometry (Metcalfe et al 1995). The slope measured in the APM
data at $16<B<19$ is $0.59$, while a linear fit to our data in this
same range of magnitudes gives a slope of $0.50\pm 0.03$. Our
measured slope is consistent with passive evolution models, which
have slopes of 0.52 and 0.51 for $q_0=0.5$ and 0 respectively, and
agrees with that of Bertin \& Dennefeld (1996), who surveyed 145
square degrees using individually calibrated Schmidt plates. We
have normalized the flat-universe passive-evolution model with a
least-squares fit to our data. This normalization is equivalent to
using a Schechter luminosity function with $b_J^* = -19.50 + 5
log(h)$ and $\alpha = -0.97$, as measured by Loveday et al.\ (1992),
but with $\phi^* = 2.02 \times 10^{-2} h^{3} Mpc^{-3}$, a normalization
that is a factor of 1.44 times higher than that measured in the
Stromlo-APM survey (Loveday et al.\ 1992.) High normalization models
have been proposed to reduce the excess of faint blue galaxies that
has been seen in deep photometric surveys (for a review, see Metcalfe
et al.\ 1996), and to fit the WFPC2 Medium Deep Survey results
(Glazebrook et al.\ 1995; Driver et al.\ 1995). The sensitivity of
our survey to low surface brightness galaxies, and the effects of
clustering on the errors of the number counts will be discussed
elsewhere (Gardner et al., in preparation).

\section{Conclusions}

We have presented bright galaxy number counts measured with linear
detectors in the $B$, $V$, $I$, and $K$ bands in two fields totalling
nearly 10 square degrees. All of our measurements are consistent
with passive evolution models. Our counts do not exhibit the steep
slope measured in other surveys, either in the $K-$ or $B-$ bands,
and so do not support earlier interpretations that required a large
amount of luminosity evolution at low redshift. We also do not find
evidence of a large underdensity in the local universe, unless it
is a phenomenon occuring exclusively in the South Galactic Pole
region. (Both of our fields are North of the Galactic Plane.)  Our
data are consistent with the conclusions of Metcalfe et al (1995)
that the steep slope measured previously in the bright $B-$band
number counts is most likely due to systematic errors in the
non-linear photometry of photographic plates. Our $B-$band counts
support the high normalization models, based upon a local $\phi^*$
approximately 1.4 times higher than that measured by Loveday et
al.\ (1992). Our $K-$band counts are consistent with previous
measurements, and have reached a fair sample of the universe in
the region where evolution and K-corrections are well understood.

\section*{Acknowledgments}

We would like to thank Carlton Baugh, Tom Shanks, Nigel Metcalfe
and Simon White for useful discussions. We acknowledge generous
allocations of time at the Kitt Peak National Observatory. This
work was supported by a PPARC rolling grant for Extragalactic
Astronomy and Cosmology at Durham.

\end{document}